*Photoinduced metastable cation disorder in metal halide double perovskites*

Shunran Li[1,2], Burak Guzelturk[3], Conrad A. Kocoj[1,2], Donald A. Walko[3], Du Chen[1,2], Haidan Wen[3,4], Xian Xu[2,5], Xiaoming Wang[6], Bongjun Choi[7], Borui Li[1,2], Zhibo Kang[8], Cunming Liu[3], Suchismita Sarker[9], Benjamin T. Diroll[10], Xiaoyi Zhang[3], Yong Q. Cai[11], Yu He[8], Deep Jariwala[7,12], Yanfa Yan[6], Diana Y. Qiu[2,5], Peijun Guo[1,2,*]

[1]Department of Chemical & Environmental Engineering, Yale University, New Haven, CT 06520, USA
[2]Energy Sciences Institute, Yale University, West Haven, CT 06516, USA
[3]X-ray Science Division, Argonne National Laboratory, Lemont, IL 60439, USA
[4]Materials Science Division, Argonne National Laboratory, Lemont, IL 60439, USA
[5]Department of Materials Science, Yale University, New Haven, CT 06520, USA
[6]Department of Physics and Astronomy, Wright Center for Photovoltaics Innovation and Commercialization (PVIC), The University of Toledo, Toledo, OH, 43606 USA
[7]Department of Electrical and Systems Engineering, University of Pennsylvania, Philadelphia, Pennsylvania 19104, USA
[8]Department of Applied Physics, Yale University, New Haven, CT 06520, USA
[9]Cornell High Energy Synchrotron Source, Cornell University, Ithaca, NY 14853, USA
[10]Center for Nanoscale Materials, Argonne National Laboratory, Lemont, IL 60439, USA
[11]National Synchrotron Light Source II, Brookhaven National Laboratory, Upton, New York 11973, USA
[12]Department of Materials Science and Engineering, University of Pennsylvania, Philadelphia, Pennsylvania 19104, USA

*E-mail: peijun.guo@yale.edu

**Abstract**

Lead-free perovskites have emerged as environmentally benign alternatives to lead-halide counterparts for optoelectronics. Among them, the double perovskite $Cs_2AgInCl_6$ family exhibits remarkable white-light emission with proper composition engineering, enabled by strong electron-phonon coupling and the formation of self-trapped excitons (STEs). Despite these advantages, the fundamental photo- and structural dynamics governing their excited-state behavior remain poorly understood. Here, we report a long-lived metastable phase in the $Cs_2AgInCl_6$ double perovskite family and unravel this process and the concomitant electronic and structural evolution using a suite of tools including transient optical spectroscopy, time-resolved X-ray diffraction (TR-XRD) and X-ray absorption (TR-XAS). We show that the photoinduced, transient metastable phase is associated with B-site (Ag-In) disorder, which induces a dramatically reduced optical bandgap. Supported by TR-XRD and first-principles calculations, the Ag-In disorder drives the formation of Ag-rich and In-rich domains with millisecond lifetimes, with lifetimes increasing at lower temperatures. TR-XAS further reveals that photogenerated STEs oxidize $Ag^+$ to $Ag^{2+}$, facilitating this highly temporally asymmetric order-disorder transition. Our findings demonstrate a new mechanism, mediated by hole-localized STE formation, that enables prolongation of transient light-induced states to the multi-millisecond regime in double perovskites, opening possibilities to harvesting the functional properties of metastable phases of these materials.

**Main text**

Metal halide perovskites have attracted intense research interest over the past decade due to their outstanding optoelectronic properties, including tunable direct bandgaps, high optical absorption coefficients, long carrier lifetimes, and solution processability.[1, 2, 3, 4] These features enabled wide-ranging applications in photovoltaics,[5, 6] light-emitting diodes,[7, 8] and photodetectors,[9, 10] to name a few. Lead-based perovskites exemplified by (MA/FA)PbI$_3$ and (MA/FA)PbBr$_3$ have led to record-holding power conversion efficiencies in perovskite solar cells and high external quantum efficiencies in perovskite light-emitting devices.[11, 12, 13] Despite these successes, concerns about lead toxicity and long-term material instability under environmental stressors have motivated efforts to develop lead-free alternatives.[14, 15] One promising class of materials is lead-free double perovskites, which adopt the A$_2$B$^+$B$^{3+}$X$_6$ structure by substituting pairs of Pb$^{2+}$ ions with monovalent and trivalent cations.[16, 17, 18] This heterovalent substitution not only retains the desirable 3D perovskite framework but also enables a diverse selection of elements for tailored electronic and optical properties.[19, 20] Among various double perovskites, Cs$_2$AgBiBr$_6$ attracted attention in the fields of solar cells and photodetectors, whereas Cs$_2$AgInCl$_6$ and its Na-doped analog Cs$_2$(Na,Ag)InCl$_6$ have recently gained increasing interest in light-emitting applications.[21, 22]

Similar to lead-based perovskites, double perovskites also exhibit the characteristic softness of their lattices.[23] Under thermal activation, B-site cation disordering occurs in Cs$_2$AgBiBr$_6$, accompanied by a reduction in bandgap.[24] In contrast, the lattice of Cs$_2$AgInCl$_6$ is more thermally robust and shows no bandgap change up to 400 K.[21] However, Cs$_2$AgInCl$_6$ exhibits exceptionally strong electron–phonon coupling, where photoexcited carriers rapidly induce a characteristic Jahn–Teller distortion of the [AgCl$_6$] octahedra and the formation of self-trapped excitons (STEs) within a few hundred femtoseconds (fs) (**Fig. 1a**).[25, 26] This strong coupling leads to a broad, strongly Stokes-shifted photoluminescence (PL) emission spectrum spanning the entire visible range (400–800 nm, **Fig. 1b**).[21] STE-dominated carrier dynamics can be effectively described using a configuration coordinate diagram (**Fig. 1c**), where STEs are intrinsic, transient structural distortions associated with the excited state of trapped charge carriers stabilized in the soft lattice. Time-resolved PL and picosecond (ps)-to-nanosecond (ns) transient absorption (TA) spectroscopy have been utilized to investigate the excited-state dynamics of STEs, and the broadband PL and sub-bandgap photoinduced absorption (PIA) are considered a characteristic signature of STEs.[27, 28, 29] However, previous TA measurement employed a ns delay time window, which was three orders of magnitude shorter than the STE lifetime (~μs). The assignment of PIA to the formation of STEs was not well substantiated.

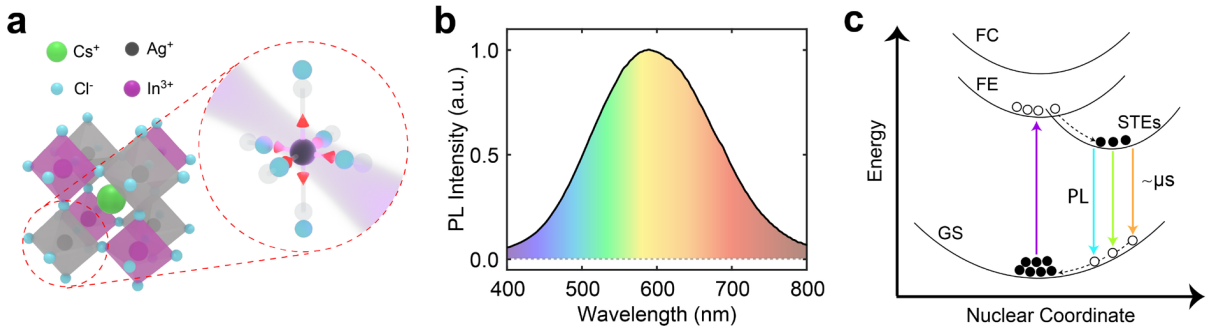

**Fig. 1 | Conventional understanding of exciton self-trapping in double perovskite $Cs_2AgInCl_6$. a** Crystal structure of $Cs_2AgInCl_6$ (CAIC), and a schematic illustration of the Jahn-Teller distortion of Ag-centered octahedra formed upon photoexcitation. **b** Photoluminescence (PL) spectrum of Na-doped $Cs_2AgInCl_6$ (CNAIC) measured at room temperature. **c** Configurational coordinate diagram showing the electronic processes after pulsed photoexcitation (FC: free carrier. FE: free exciton. STE: self-trapped exciton. GS: ground state. PL: photoluminescence).

Here, to address whether the PIA is driven by STE formation, we investigate the lifetime of the two processes with an extended time window which covers the full decay of STEs. Notably, we identified an additional, much longer long-lived process, distinct from STEs, that leads to the broadband PIA. By combining time-resolved X-ray diffraction (TR-XRD) and time-resolved X-ray absorption spectroscopy (TR-XAS), we demonstrate that this long-lived PIA originates from a reversible translocation between the $[AgCl_6]$ and $[InCl_6]$ octahedra. This translocation leads to the formation of Ag-rich and In-rich nanoscale domains and reduces the bandgap of $Cs_2AgInCl_6$, similar to the thermally induced B-site disorder reported in $Cs_2AgBiBr_6$. However, in $Cs_2AgInCl_6$ this B-site translocation is triggered by STEs and thus exhibits a highly asymmetric temporal behavior: it develops on a sub-nanosecond timescale but requires several milliseconds (ms) to relax back to the ground-state ordered phase. The photo-induced cation swapping, which is expected to show even longer lifetimes at liquid helium temperature, marks a new mechanism to the stabilization of metastable phases for the exploitation of their functional properties.[30, 31, 32, 33]

**Photoinduced broadband absorption with extended lifetimes**

Compared with $ABX_3$-type perovskites, the $A_2B^+B^{3+}X_6$-type double perovskites exhibit higher compositional complexity.[22, 34] In $Cs_2AgInCl_6$-based system, both $Ag^+$ and its substitutional $Na^+$ dopant occupy the $B^+$ site, while $In^{3+}$ with a trace amount of $Bi^{3+}$ dopants take the $B^{3+}$ sites. Previous studies showed that $Na^+$ primarily modifies the band structure and the spatial distribution of electronic wave functions, altering charge-carrier and STE properties including the bandgap and exciton binding energy.[21] In contrast, trace $Bi^{3+}$ passivates defects and enhances the emission efficiency of STEs.[21] In this work, we primarily investigate the properties of both pristine double perovskite $Cs_2AgInCl_6$ (abbreviated as CAIC; PLQY<1%) and the highly emissive $Cs_2Ag_{0.6}Na_{0.4}InCl_6$ with 0.5% Bi doping (abbreviated as CNAIC; PLQY~90%). Their static optical and structural properties are summarized in **Supplementary Note 1**.

We first examined the charge carrier dynamics in the more emissive CNAIC with in-house optical spectroscopy. As presented in **Fig. 2a**, time- and wavelength-resolved PL measurement reveals that the broadband PL emission associated with STEs decays over several microseconds (μs). The PL lifetime exhibits negligible wavelength dependence (**Fig. S4**), indicating that the broad PL emission arises from a single STE species rather than multiple emissive STE states.[35] Temperature-dependent PL data shown in **Fig. 2b** reveal modestly enhanced PL lifetimes achieved at lower temperatures, although still in the μs range. This behavior is expected, as the high PLQY of CNAIC suggests that radiative recombination is the dominant carrier relaxation pathway, while temperature, which primarily influences nonradiative channels, does not strongly influence the radiative rates and hence the PL lifetime.

Following time-resolved PL measurements, we then probed the carrier dynamics using TA spectroscopy measured in transmittance configuration. Generally, in light-emitting materials with large Stokes shifts, a transient enhanced transmission within the PL spectral regime is expected typically due to stimulated emission.[36] However, previous studies on metal-halide perovskite-based STE emitters have reported broadband photoinduced absorption (PIA) in the visible range upon photoexcitation.[21, 25] The broad PIA was attributed to excited-state transitions of STEs to higher-lying energy levels in the conduction band. Nevertheless, those previous studies explored time windows up to a few ns using mechanical delay lines, leaving the μs-long decay dynamics of STE-related absorption uncharted. By extending the TA measurement window to several tens of μs using an electronically delayed probe, which, in principle, is sufficient to capture the complete recombination dynamics of STEs in correlation with time-resolved PL data, we found, surprisingly, that the broadband PIA persists far longer than the STE PL lifetime.

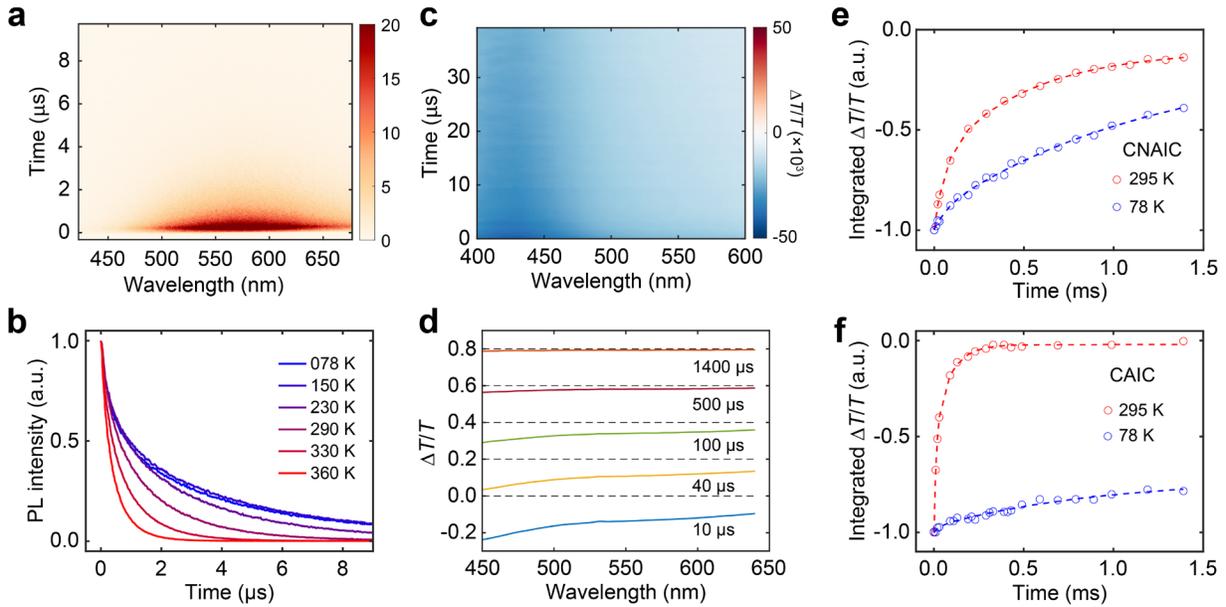

**Fig. 2 | Kinetics discrepancy between photoluminescence and transient absorption signals. a** Time- and wavelength-resolved PL, and **b** Wavelength-integrated PL kinetics of CNAIC from room temperature down to 78 K. **c** Transient $\Delta T/T$ spectral map of CNAIC in the 0~40 μs delay time window. **d** Transient

$\Delta T/T$ spectra of CNAIC at selected delay times in the 0~1400 μs window. The neighboring curves are offset by 0.2 for clarity. **e-f** $\Delta T/T$ kinetics of CNAIC in **e** and CAIC in **f**, respectively, both at 295 K and 78 K, spectrally integrated from 500 nm to 600 nm.

As shown in **Fig. 2c**, μs-TA spectroscopy data shows a strong residual PIA signal persisting beyond 35 μs. In addition to the long-lived PIA, a non-zero signal is also observed before time zero (**Fig. S5**), indicating that the PIA decay exceeds the pump pulse interval of 1 ms. To capture the full decay of this long-lived PIA component, we reduced the pump repetition rate from 1 kHz to 100 Hz and focused on delay times beyond the complete decay of STEs (*i.e.*, several μs after the pump excitation). As revealed in **Fig. 2d**, the non-STE-related broadband PIA decays within *several hundred* μs, and its broadband spectral shape remains nearly unchanged throughout the decay process (**Fig. S6**). The extendedly long-lived PIA indicates the presence of a previously unidentified state of matter – distinct from STEs – that may be activated upon photoexcitation. In addition, the invariance of the PIA spectral profile implies that it is governed by a single, well-defined mechanism.

**Fig. 2e** presents the normalized, wavelength-integrated transient $\Delta T/T$ of CNAIC at room temperature and 78 K, revealing an elongated lifetime at lower temperature. Similar measurements performed on undoped CAIC likewise exhibit longer-lived PIA signals (**Fig. 2f**). Notably, the PIA in CAIC shows a stronger temperature dependence than that in CNAIC. As detailed in **Supplementary Note 2**, we further investigated the influence of compositional engineering on the PIA in CAIC-based double perovskites and found that the PIA is an intrinsic characteristic of the CAIC system, irrespective of $Na^+$ and/or $Bi^{3+}$ incorporation, which only moderately modulates the PIA decay dynamics.

Combining the PL and TA results, meanwhile considering the large difference in the STE emission efficiency of CAIC and CNAIC (~90% in the latter), we hypothesize that the long-lived PIA originates from a mechanism distinct from the excited STE state. In general, when band-structure modifications induced by electronic excitation can be excluded, a plausible origin of a long-lived TA signal is photothermal effect.[37, 38, 39] However, because both CNAIC and CAIC possess bandgaps in the ultraviolet region (**Fig. S2**),[21] the temperature-induced bandgap shift – *i.e.*, changes in the imaginary part of the complex refractive index – should not result in significant broadband absorption in the visible range. The potential contribution of the real component of the refractive index, which might alter the transmittance through changes in interfacial reflectance, was thoroughly examined and excluded in **Supplementary Note 3**.

We note that previous studies on the $Cs_2AgBiBr_6$ analogue revealed a thermally induced bandgap reduction.[24, 40] Importantly, the bandgap narrowing did not originate from temperature-induced lattice expansion but rather from temperature-induced cation disordering. Given the structural similarity between CAIC and $Cs_2AgBiBr_6$, we hypothesize that a similar lattice reconfiguration, corresponding to a metastable phase, occurs in CAIC-based system, which can substantially reduce the bandgap from the ultraviolet to the near-infrared region. However, unlike that in $Cs_2AgBiBr_6$, the lattice reconfiguration in CAIC is not thermally triggered (the material is thermally stable from 77 K up to 400 K without a change in bandgap and PL spectra),[21] but instead driven by photo-excited STEs.

**The lattice dynamics correlated to STEs**

Both the formation of STEs and the hypothesized metastable phase involve substantial lattice reorganization. Going beyond optical spectroscopic measurements of photoexcited charge-carrier dynamics, we performed TR-XRD at the Advanced Photon Source (APS) to directly probe transient lattice dynamics of CNAIC with X-ray pulses following pulsed photoexcitation at 360 nm (**Fig. 3a**, see Methods). The (222) plane, as the natural habit plane for solution-grown, multi-millimeter-sized CNAIC and CAIC single crystals, can be readily accessed in a standard $\theta$–$2\theta$ geometry in the TR-XRD experiments. The X-ray detector was electronically gated to selectively capture X-ray pulses synchronized with the pump laser pulses at the same repetition rate. The pump-probe delay time was continuously varied with a delay generator, and the time resolution was ~150 ps determined by the X-ray pulse width. Given the crystal orientation, the scans correspond naturally to rocking curves of the (222) plane. Rocking curves of other in-plane diffraction peaks were measured by varying the azimuthal orientation of the sample, as described in **Supplementary Note 5**.

 **Fig. 3b** shows the X-ray diffraction beam profiles corresponding to the (222), (220), and (400) planes under laser-off conditions. Despite the compositional complexity of CNAIC, the crystal quality is high, as each diffraction spot is extremely sharp—nearly limited to a single detector pixel in width. The elongated shape of the spots arises from the geometry of the X-ray source at the beamline. **Fig. S14** shows the rocking curves of the three peaks of CNAIC measured under laser-off conditions. All three peaks exhibit narrow full widths at half maximum of only ~0.01 Å$^{-1}$, supporting the exceptional crystalline quality of CNAIC. Inelastic X-ray scattering data (**Fig. S10-12**, and **Supplementary Note 4**) reveals almost identical acoustic phonon branches between CNAIC and CNIC, suggesting excellent long-wavelength structural uniformity in the doped system, consistent with the sharp X-ray diffraction peak.

 We first employed a 1-kHz laser repetition rate to probe the lattice dynamics within several μs delay-time window after photoexcitation. Based on the TA data, under these conditions the metastable phase could not fully decay before the arrival of the next laser pulse; however, within the μs timescale of interest, the ms-long decay component can be considered as a constant, pump-induced background. **Fig. 3c** presents the temporal evolution of the total integrated intensity of the (222), (220), and (400) diffraction peaks following 360-nm pump excitation. All three diffraction peaks exhibit an instantaneous increase in intensity after photoexcitation, followed by a gradual decay. Single-exponential fitting of the decay dynamics yields time constants on the order of several μs, comparable to the STE lifetime. These data indicate that STE formation enhances the total X-ray diffraction intensity, a phenomenon previously reported for photoexcited perovskites.[41]

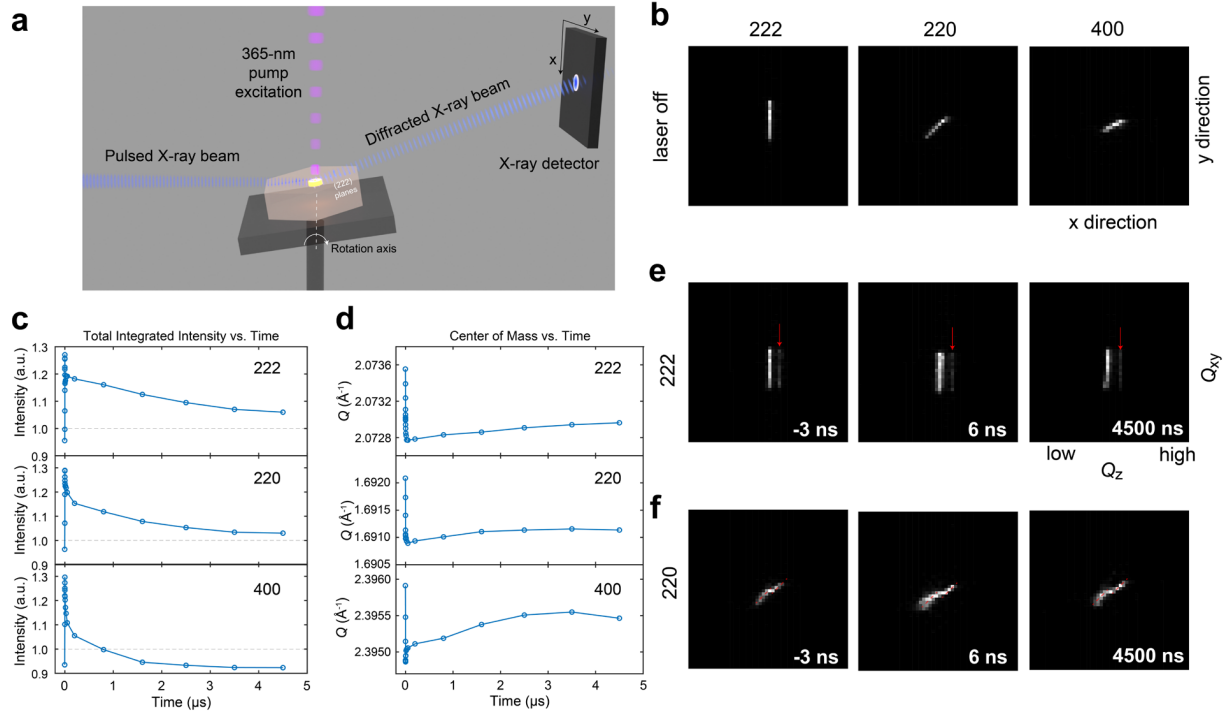

**Fig. 3 | TR-XRD response during the first several μs delay-time window (1 kHz laser pump rate). a** Schematic of the TR-XRD experiments. The incident X-ray beam is aligned horizontally, while the laser is directed vertically from above. The sample rotation axis is orthogonal to both the laser and the incident X-ray beam. **b** X-ray diffraction beam profiles of the (222), (220), and (400) planes of CNAIC measured under *laser-off* conditions. The x and y direction here correspond to the x and y direction shown in **a**. **c** Temporal evolution of integrated diffraction intensity after photoexcitation, showing an instantaneous rise followed by μs-scale decay characteristics consistent with STE relaxation. **d** Temporal evolution of the peak center-of-mass positions after photoexcitation, showing lattice expansion induced by STEs. **e** (222) diffraction profiles collected on the X-ray 2D detector under laser-on conditions at –3 ns, 6 ns, and 4500 ns delay times. A secondary peak appears on the high-$Q$ side under illumination highlighted with red arrows, signifying a contracted, photoinduced metastable phase. **f** (220) diffraction profiles collected on the X-ray 2D detector under laser-on conditions at –3 ns, 6 ns, and 4500 ns delay times. The photoinduced phase segregation is highlighted by the red dash lines. All the images in **Fig. 3** are plotted in linear scale.

Focusing on the early time response (**Fig. S17**) and comparing the diffraction intensity at 3 ns before laser excitation with that under dark conditions reveals a slightly weaker signal at –3 ns, suggesting that the presence of the metastable phase – after the STE and photothermal contributions have largely vanished – reduces the overall diffraction intensity. **Fig. 3d** presents the temporal evolution of the diffraction peak center-of-mass positions following photoexcitation. A transient decrease in $Q$ is observed immediately after photoexcitation, indicating a prompt lattice expansion, followed by a gradual recovery. The recovery dynamics are well-described by a single-exponential decay similar to their intensity decay, demonstrating that STE formation induces a transient lattice expansion.

To further elucidate the signature of the metastable phase in the TR-XRD data, we closely examined the diffraction profiles on the detector, as displayed in **Fig. 3e** and **Fig. S18**, using linear and logarithmic color scales, respectively. The measurements correspond to –3 ns (when STEs generated by the previous laser pulse have fully decayed, leaving only the residual metastable phase), 6 ns (when both STEs and metastable phases coexist before their decay), 4500 ns (when most STEs have decayed while the metastable phase remains largely intact), and under laser-off conditions (where neither STEs nor metastable phases are present). Under dark conditions, the (222) peak appears as a single-pixel-width narrow line, whereas under illumination, a weaker secondary diffraction spot consistently emerges on the right side of the main spot. As discussed in **Supplementary Note 6**, the peak on the right side corresponds to a larger $Q$ value, corresponding to a smaller lattice constant. Notably, this additional peak persists even at a negative delay time of -3 ns. Given the laser repetition rate of 1 kHz, this observation reconfirms that the structural response induced by the previous laser pulse persists for over 1 ms before fully relaxing, consistent with the long-lived PIA shown in **Fig. 2**. The emergence of this distinct, photoinduced peak thus provides further evidence for the existence of a photo-induced metastable phase.

Moreover, the diffraction peak corresponding to the metastable phase exhibits nearly identical shape and position on the detector as that of the original double perovskite, suggesting that it retains the same space group ($Fm\bar{3}m$) and differs only in lattice constant. Specifically, the photoinduced peak is displaced by approximately four pixels along the X direction from the main peak. Based on the linear correlation between detector X-pixel position and scattering vector $Q$ established in **Fig. S16**, it corresponds to a $Q$ shift of 0.0045 Å$^{-1}$, and the lattice constant of the metastable phase is estimated to be 10.477 Å, slightly smaller than that of the parent double perovskite (10.500 Å) by 0.023 Å.

**Fig. 3f** shows the diffraction peak profile of the (220) reflection at the same pump–probe delay as in **Fig. 3e**. A similar peak splitting caused by phase segregation is observed, exhibiting consistent temporal characteristics. The diffraction spot of the (400) peak likewise exhibits photoinduced distortion (**Fig. S19**), although the segregation is less pronounced compared to the (222) and (220) reflections. Therefore, this metastable phase accompanied by the long-lived PIA can be identified as a photoinduced phase segregation in double perovskites.

**The lattice dynamics correlated to the B-site disordering**

Our TA and TR-XRD results both indicate that the relaxation of the photoinduced phase recovers over hundreds of μs to single-digit ms, exceeding the continuously variable pump-probe delay range of the APS beamline. To capture the slower dynamics, we reduced the pump laser repetition rate to 167 Hz to ensure complete relaxation of the metastable phase between adjacent pump pulses. Furthermore, by adjusting the X-ray detector gating pulse (note that X-ray probe pulses are thousands of times higher in repetition rate than the optical pump), we achieved extended, but discrete, pump–probe time delays.

**Fig. 4a** shows the (222) diffraction peak profiles under laser-on and laser-off conditions at a delay time of 147 μs. At this delay time, when STEs are fully relaxed, the metastable phase leads to a reduction in intensity at the peak center and an increase on the shoulders. **Fig. 4b** displays the temporal evolution of Bragg peak intensity on the high-$Q$ side, highlighting the contrasts between laser-on and laser-off conditions. Upon photoexcitation, a pronounced intensity enhancement appears in this region, which gradually decays over several-hundred μs. This decay characteristic is consistent with the PIA observed in TA experiments.

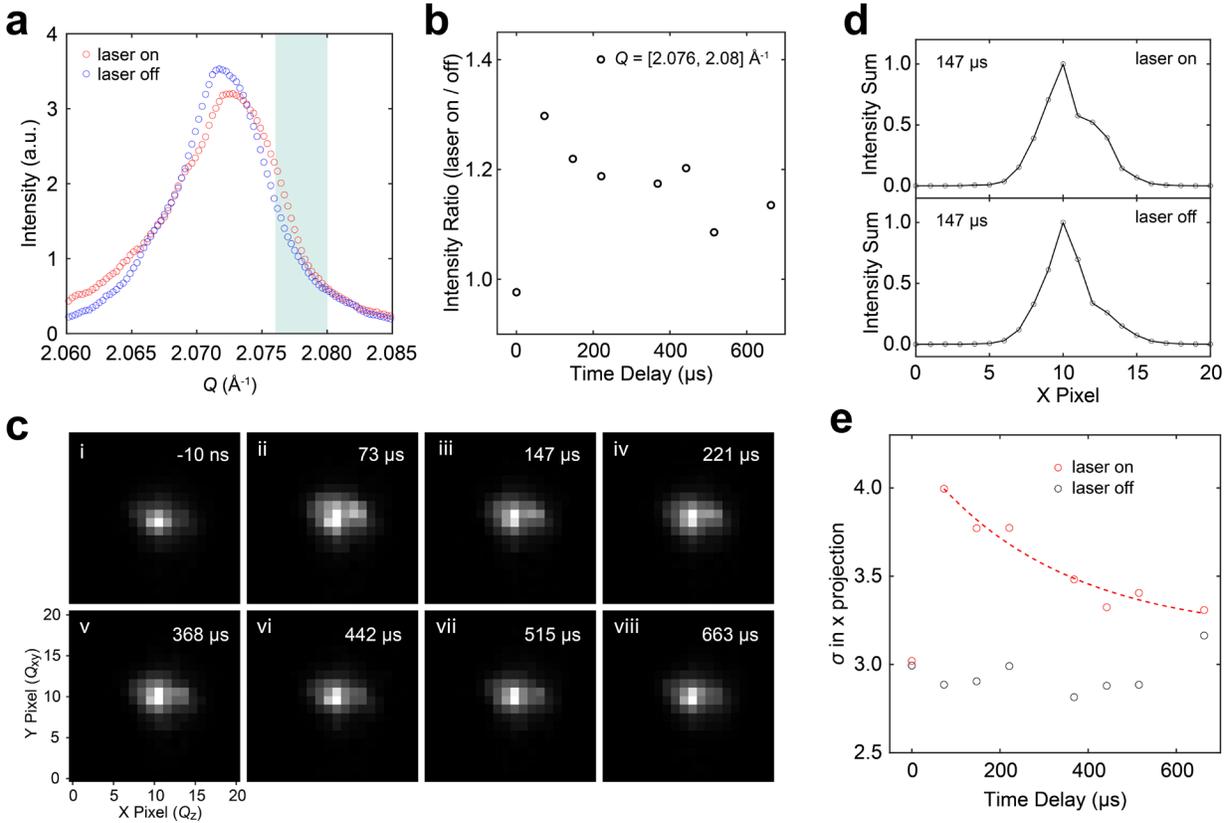

**Fig. 4 | TR-XRD response in the ms time range. a** Rocking curve of the (222) plane without pulsed laser excitation (blue) and with pulsed laser excitation captured at a delay time of 147 μs (red). **b** Temporal evolution of Bragg peak intensity upon laser excitation, integrated from 2.076 to 2.08 Å$^{-1}$. **c** Temporal evolution of the diffraction images on the X-ray detector, which is obtained by integrating diffraction spots within the $Q$ range of 2.060 to 2.065 Å$^{-1}$. **d** X-directional diffraction intensity profile with/without (upper/lower panel) laser excitation, obtained by integrating the 2D diffraction pattern along the y-axis. **e** Time evolution of the standard deviation (σ) along the x-direction, extracted from Gaussian fits to the diffraction intensity profiles.

We further compared the diffraction spot profiles at various delay times. As shown in the first panel of **Fig. 4c**, the diffraction spot before time-zero appeared symmetric, confirming that the metastable phase fully decayed before the arrival of the next laser pulse. Note that the spot shape is more circular, owing to improved beam coherence and change in focusing optics after the APS upgrade. Nevertheless, after time zero, an intensity enhancement appears on the right side of the main spot, corresponding to the formation of the metastable phase as seen in **Fig. 3e-g**.

Since the X coordinate of the detector correlates with $Q$ for the (222) reflection whereas the Y coordinate does not convey information about lattice constant, we integrated the raw 2D diffraction data along the Y direction to obtain the intensity distribution along the $2\theta$ direction (**Fig. 4d**). Compared with the laser-off condition, the laser-on curve exhibits a clear intensity enhancement on the right side associated with the metastable phase. **Fig. 4e** shows the temporal evolution of the standard deviation of the diffraction-spot distribution along the $2\theta$ direction under laser-on and laser-off conditions. The standard deviation increases upon photoexcitation and gradually decays over several hundred μs, whereas it exhibits only random fluctuations in the laser-off state. It should be noted that the diffraction signals of the original and metastable phases overlap substantially, rendering peak deconvolution and extraction of independent dynamics of the metastable phase experimentally less practical.

**The role of STEs in the metastable phase formation and recovery**

By combining TA and TR-XRD measurements, we confirmed the existence of a photoinduced metastable phase segregation in CAIC and CNAIC, evidenced by the optical PIA signal and the XRD peak splitting. In perovskite materials, ionic substitution typically occurs between isovalent ions.[42, 43, 44] For instance, in MAPbI$_x$Br$_{1-x}$ ($x$=0~1), phase segregation under illumination leads to the formation of I-rich and Br-rich regions.[45, 46] Considering the structure of Cs$_2$B$^+$B$^{3+}$Cl$_6$, the exchange between B$^+$ and B$^{3+}$ cations is a plausible process responsible for the observed phase segregation, as detailed below.

We illustrate the photodynamic cation swapping mechanism in **Fig. 5a**. The swapping process induces phase segregation, which underpins the metastable phase in CAIC and its derivative CNAIC. In the ground state, the Ag- and In-centered octahedra are ordered. Upon photoexcitation, electron–hole pairs are generated, and within several-hundred fs, the hole oxidizes Ag$^+$ into Ag$^{2+}$. Accompanying this, the originally symmetric octahedron undergoes a Jahn–Teller distortion and, through strong electron–phonon coupling, forms a STE. Immediately afterwards (within sub-ns), Ag$^{2+}$ undergoes a site-exchange with a neighboring In$^{3+}$. This cationic swapping results in the formation of locally Ag- and In-enriched domains, with the Ag domains exhibiting a significantly reduced bandgap. Together with the STEs, the aggregated domains contribute to the observed broadband PIA. Following this, the STEs recombine over a few μs, emitting broadband PL, while Ag$^{2+}$ is reduced back to Ag$^+$. The size-mismatched Ag$^+$ and In$^{3+}$ then gradually return to their original, ordered lattice positions over several ms, driven by thermal fluctuations with a large energy barrier, completing the full photodynamic cycle.

As discussed in **Supplementary Note 8**, DFT calculations were performed to determine the lattice constants of In-rich and Ag-rich domains. Compared with the Ag–In ordered double perovskite structure, the In-rich domain exhibits a smaller lattice constant, whereas the Ag-rich domain shows a larger one. This indicates that the photoinduced diffraction peak observed in TR-XRD corresponds to the In-rich domain, despite that Ag-rich and In-rich domains coexist. This is rationalized by the fact that the measured CNAIC sample contains both Ag and Na cations at the B$^+$ sites; under the condition of B$^+$-rich and B$^{3+}$-rich domains, the multicomponent (Ag–Na) B$^+$- composition reduces the crystallinity and thus the XRD intensity of the $B^+$-rich domains, but not the $B^{3+}$-rich domains. Therefore, only the diffraction from the compositionally purer In-rich domain was experimentally observed. Furthermore, as shown in **Supplementary Note 9**, a single Ag–In exchange in a 2×2×2 CAIC supercell is sufficient to reduce the bandgap by at least 1.19 eV, which is consistent with the broadband PIA feature across the entire visible range observed in TA spectroscopy (**Fig. 2**).

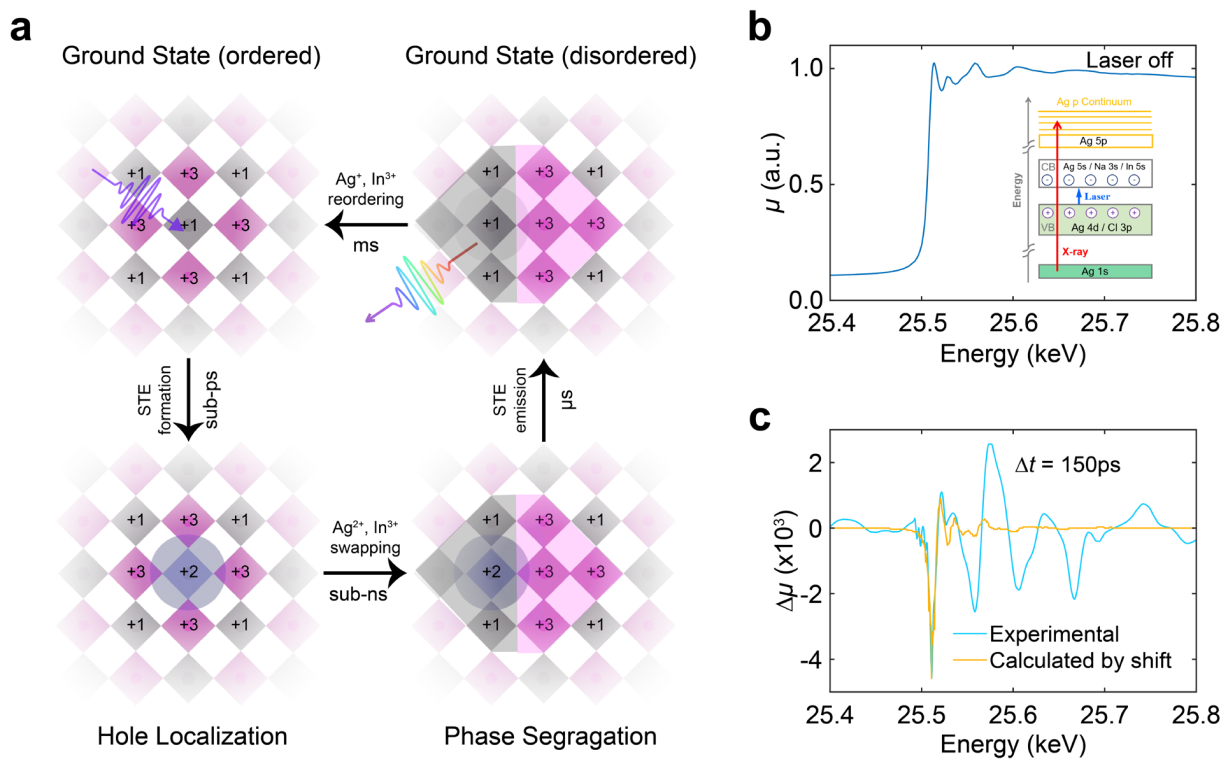

**Fig. 5 | Schematic depiction of the photo-driven lattice dynamics, and TR-XAS results. a** Schematic drawing of the photo-induced structural rearrangements in CAIC, ranging from the sub-ps to the ms timescales. **b** XAS captured at the Ag K-edge in the ground-state (dark) condition. **c** Differential XAS spectra due to photoexcitation (pump-probe delay time is 150 ps). The cyan curve shows the measured spectrum from TR-XAS experiments; the orange curve shows the differential XAS spectrum obtained by manually shifting the dark spectrum in **b** by 35 meV.

Importantly, the process exhibits an asymmetric temporal behavior: it forms within a sub-ns timescale but decays on the ms timescale. In contrast, photoinduced phase transitions in lead halide perovskites—whether driven by photothermal effects or photogenerated charge carriers—do not show such an asymmetric response.[47] Instead, the forward and recovery phase-change processes in both 3D and 2D metal-halide perovskites occur within tens to hundreds of ns.[47, 48] We attribute this pronounced temporal asymmetry observed here to the unique nature of STEs. Specifically, in CAIC and CNAIC, the electron and hole wavefunctions of the STEs are asymmetric: the hole is highly localized on $Ag^+$, whereas the electron is delocalized over several unit cells.[21] Therefore, the locally STE-associated $Ag^+$ is oxidized into $Ag^{2+}$. This oxidation process is critical, as cation swapping processes are more favorable between more isovalent ions; the smaller charge difference between $Ag^{2+}$ and $In^{3+}$ lowers the energy barrier for cation exchange, when compared with $Ag^+$-$In^{3+}$ swapping. Additionally, $Ag^+$ (1.15 Å) has a larger ionic radius than $In^{3+}$ (0.8 Å); upon oxidation to $Ag^{2+}$, its radius decreases, reducing the energy barrier for swapping. After STEs recombine, $Ag^{2+}$ reverts back to $Ag^+$, and the large charge and ionic radius differences between $Ag^+$ and $In^{3+}$ increase the energy barrier for relaxation of the metastable phase, giving rise to the asymmetric relaxation behavior.

The influence of ionic radii difference on energy barrier provides a compelling explanation for the contrasting $B^+$–$B^{3+}$ cation exchange behaviors observed in $Cs_2AgBiBr_6$ and CAIC. In $Cs_2AgBiBr_6$, diffuse scattering signals have been observed in single crystals, which is a hallmark of a new type of short-range B-site arrangement induced by $Ag^+$–$Bi^{3+}$ exchange, in line with its thermochromic behavior.[49, 50] In contrast, CAIC does not exhibit thermochromic behavior (**Fig. S9**), and our high dynamic-range reciprocal space mapping of CAIC and CNAIC, across a range of temperatures, reveal a consistently high degree of crystallographic order, with no evidence of diffuse scattering (**Fig. S24-25**). This discrepancy between $Cs_2AgBiBr_6$ and CAIC can be partly explained by the different ionic radii of $Ag^+$ (1.15 Å), $Bi^{3+}$ (1.03 Å), and $In^{3+}$ (0.8 Å): $Ag^+$ is 43.7% larger than $In^{3+}$ but only 11.6% larger than $Bi^{3+}$. The larger mismatch between $Ag^+$ and $In^{3+}$ radii leads to a higher energy barrier for their exchange, thereby suppressing thermal disorder in CAIC.

The photo-driven redox process involving $Ag^+$ is central to the complete photodynamic pathway described above. However, $Ag^{2+}$ is an unstable, rare oxidation state. So far, the oxidation of $Ag^+$ by STEs has been proposed only through theoretical calculations and lacks direct experimental evidence.[21] To experimentally verify the STE-induced oxidation of $Ag^+$, we directly probe the electronic structure of Ag cation in CNAIC nanocrystals with TR-XAS. The TR-XAS measurements were performed under 355-nm, pulsed laser excitation (~10 ps pulse width) and ~150 ps X-ray pulses from the APS as the probe. **Fig. 5b** displays the Ag K-edge X-ray absorption near edge structure (XANES) spectrum of CNAIC acquired without laser excitation. This spectrum features a prominent white-line transition at 25.5232 keV, attributed to the dipole allowed 1s→5p transition.[51] With a 150-ps time delay after the 355-nm laser excitation, the XANES spectrum shows a clear difference signal: a negative feature observed at 25.5205 keV indicates a shift of the absorption edge toward higher energy as shown in **Fig. 5c**. By manually shifting the absorption spectrum and calculating the difference, we further determined this blueshift to be approximately 35 meV. Importantly, thermal effects on XAS are generally minimal and tend to cause redshifts (toward lower energies), instead of blueshifts.[52] Therefore, this observed blueshift provides direct experimental evidence that STEs induce the oxidation of $Ag^+$ into $Ag^{2+}$.

**Discussion**

By combining optical spectroscopy and synchrotron-XRD and XAS experiments, we reveal that in addition to STEs, the highly emissive double perovskites CAIC and CNAIC exhibit a photoinduced, metastable B-site cationic reorganization. This process leads to phase segregation into Ag-rich and In-rich domains, leading to a significantly reduced electronic bandgap of the material. Remarkably, this metastable phase persists for several to tens of ms, far exceeding the STE lifetimes, highlighting a distinct, long-lived contribution to these materials' photodynamics. The B-site disorder mechanism unveiled from this study may be more broadly applicable across double perovskites, where similar lattice-driven bandgap narrowing could explain the widely observed thermochromic effects.[53, 54, 55] Moreover, the metastable nature of the segregated phase allows it to accumulate under continuous excitation, particularly approaching liquid-helium temperatures and potentially accounting for anomalous optical phenomena such as high-order Raman scattering.[56] While this bandgap reduction can hinder these materials for obtaining optical gain and stimulated emission, it offers a potential advantage for photodetection. We point out that sustained excitation with sub-bandgap light can enhance photo-response through increased carrier generation. This concept has already been successfully applied in detecting weak infrared signals, suggesting a new strategy for optimizing lead-free perovskite photodetectors.[57, 58]

## Methods

### Single crystal growth

Caesium chloride (CsCl, 99.99%), silver chloride (AgCl, 99%), sodium chloride (NaCl, 99.99%), anhydrous indium chloride ($InCl_3$, 99.99%), anhydrous bismuth chloride ($BiCl_3$, 99.99%) were purchased from Alfa Aesar without further purification. 1 mmol anhydrous $InCl_3$, 0.005 mmol anhydrous $BiCl_3$ and 2 mmol CsCl were first dissolved in 10 ml of a 10-M HCl solution in a 25-ml Teflon autoclave. Then 0.6 mmol of AgCl and 0.4 mmol of NaCl were added and the solution was heated at 180 °C for 12 h in a stainless-steel Parr autoclave for CNAIC (1 mmol AgCl was used for CAIC without any NaCl). The solution was then steadily cooled to 50 °C at a speed of 1 °C·h$^{-1}$. The as-prepared crystals were then filtered out, washed with isopropanol and dried in a furnace at 60 °C.

### Nanocrystal synthesis

Bismuth acetate (99.99%), silver acetate (99.99%), indium acetate (99.99%), cesium carbonate (99.9%), sodium acetate (99.0%), dioctylether (99%), oleylamine (70%), oleic acid (90%), and benzoyl chloride (99.0%) were obtained from Sigma-Aldrich. Chemicals were used as received. Silver acetate was stored under foil. Synthesis of Bi-doped $Cs_2Ag_{1-x}Na_xInCl_6$ nanocrystals followed literature procedures[59] with a scale-up by a factor of four. The value of $x$ in the formula was 0.4. (i.e., 40% Na). Briefly, 1 mmol cesium carbonate, indium acetate, 0.6 mmol silver acetate, 0.4 mmol sodium acetate, and 0.005 mmol bismuth acetate were added to a three-neck flask held under an air atmosphere with 16 mL dioctyl ether, 2.4 mL oleylamine, and 4.8 mL oleic acid, stirring vigorously. The reaction was heated to 115 ºC, held 5 minutes, then heated to 145 ºC. Separately, dioctylether was degassed by sparging with nitrogen. Once the reaction reached 145 ºC, an injection solution of 0.8 mL benzoyl chloride and 2.4 mL degassed dioctylether was rapidly injected. The reaction vessel is immediately cooled with an ice bath and the reaction contents centrifuged at 4500 rpm for 5 minutes, discarding the supernatant. The remaining pellet was redispersed in hexanes and precipitated with ethyl acetate (1:2 v:v) and centrifuged again 5 minutes at 6000 rpm. The solid pellet was redispersed in hexanes for further studies. Transmission electron microscopy imaging was performed using a JEOL 2100F TEM operated at 200 keV.

### Ultrafast ps-ns pump-probe TA spectroscopy

Femtosecond (fs) pump pulses were generated using an optical parametric amplifier (OPA, Orpheus-F), driven by a Pharos fs laser amplifier operating at a 2-kHz repetition rate. The amplifier produced 1030 nm fundamental pulses with a 170-fs pulse width, of which 0.5 mJ per pulse was used to power the OPA. The OPA outputs 740 nm pulses, which were frequency-doubled using a BBO crystal to produce 370 nm pump pulses. The pump beam was focused on the sample using a 250 mm focal length $CaF_2$ lens, producing a ~1 mm diameter spot. Supercontinuum probe pulses were generated by focusing a small portion of the 1030 nm output onto a 4-mm-thick sapphire window and delayed using a retroreflector. The pump and probe pulses operated at 1 kHz and 2 kHz, respectively. Probe spectra were acquired at 2 kHz using a USB spectrometer (Avantes).

**Longer delay-time pump-probe TA spectroscopy**

The pump pulses were identical to that used in ultrafast pump-probe spectroscopy. For delay times up to 50 μs, the pump and probe lasers were operated at 1 kHz and 2 kHz, respectively; the probe laser (Leukos Disco-UV) was electrically triggered by a digital delay generator (Stanford Research, DG645). For delay time window up to several ms, the pump repetition rate was reduced to 100 Hz using a 10% duty-cycle optical chopper (Thorlabs MC2000B), enabling full recovery of the sample between adjacent pump pulses. The probe was operated at 200 Hz. Pump-probe delay was achieved using a digital delay generator. The samples were mounted in a liquid-nitrogen cryostat (Lake Shore VPF-100) with a vacuum level better than $1\times10^{-4}$ Torr.

**Time-resolved photoluminescence**

The PL excitation pulses were identical to the pump pulses used in the pump-probe spectroscopy experiments. The samples were mounted in a closed-cycle liquid helium cryostat (ARS) with a vacuum level higher than $1\times10^{-5}$ Torr. The PL photons were steered into a spectrograph, which is equipped with an EMCCD camera for time-integrated spectral acquisition, and a streak camera (Hamamatsu) for time- and wavelength-resolved measurements.

**Time-resolved X-ray diffraction (TR-XRD)**

TR-XRD measurements were performed at the Beamline 7-ID-C of the Advanced Photon Source (APS) at Argonne National Laboratory (ANL). A Ti:sapphire laser amplifier's output (Legend, Coherent Inc with 1 kHz repetition rate, 70 fs pulse width and 800 nm) was fed into an optical parametric amplifier (OPA) system (Opera Solo, Coherent Inc). The fourth harmonic of the OPA signal output (360 nm, 1 kHz, 70 fs) was then routed to the sample surface. The sample was positioned at the center of rotation of a six-circle Huber diffractometer. Pump pulses at 360-nm wavelength impinged onto the sample surface close to the surface-normal. The size of the elliptical laser spot was 220 by 200 $\mu$m in diameters. Hybrid pixel photon-counting area detectors (Eiger2 500k and Pilatus 100k, Dectris Inc.) were used to collect the X-ray diffraction from the sample. The area detector was electronically gated to single X-ray pulses matching the repetition rate of the laser pulses. Experiments took place during the 24-bunch mode operation of the APS with monochromatic X-ray energy at 10 keV and pulse width of ~80 ps in full-width at half maximum before the upgrade and ~150 ps after the upgrade. The time delay between the laser pump and X-ray probe pulses were continuously varied by digital delay generator (DG 645, Stanford Research Systems) for data in Fig. 3, and varied in steps by moving the gating time of the X-ray detector for data in Fig. 4.

**Time-resolved X-ray absorption spectroscopy (TR-XAS)**

The Ag K-edge TR-XAS spectra of 1.0 mM $Cs_2Ag_{0.6}Na_{0.4}InCl_6$ nanocrystals dissolved in hexane were collected at the beamline 25-ID-E of APS at ANL. Detailed descriptions of the experimental setup and data acquisition are provided in prior publications.[60, 61] The sample was photoexcited using a 355 nm laser (~10 ps pulse duration), generated from the third harmonic output of a Duetto laser (Time-Bandwidth), operating at a repetition rate of 54.56 kHz. The X-ray probe pulses with

~150 ps pulse width were derived from the 48-bunch timing mode of the APS, which delivers evenly spaced X-ray pulses at 13.03 MHz with 76.7 ns intervals. The pump and probe beams intersected at a free-flowing liquid jet (~500 µm in diameter) enclosed in a nitrogen atmosphere. The X-ray probe pulse synchronized with the laser pump was delayed by 150 ps relative to the pump.

The laser and X-ray spot sizes (H×V in µm×µm) at the sample position were ~600×80 and 450×30, respectively. Fluorescence signals from Ag were collected at a 90º angle relative to the incident X-ray beam using two avalanche photodiodes (APDs) positioned on either side. A customized combination of Soller slits and a 25-µm thick Rh filter, which was designed to match the sample chamber geometry and detector distance, was placed between the sample jet and the APDs. Sample integrity was monitored periodically via UV-vis spectroscopy, so that the solution was refreshed once any sign of degradation was detected. The APD signals were digitized using a high-speed analyzer card (Agilent), triggered by a 54.56 kHz TTL pulse synchronized with the laser. The card recorded the X-ray fluorescence with 1 ns time resolution per point and averaged signals over a 4-second integration window. All individual X-ray pulses occurring between two laser pulses were captured. To minimize systematic error from pulse-to-pulse intensity fluctuations, the fluorescence signal from the sample was normalized to a reference signal obtained by measuring the air scattering intensity from the X-ray beam upstream of the sample. The laser-off spectrum was obtained by averaging the fluorescence signals from six X-ray probe pulses recorded prior to photoexcitation, representing the ground-state absorption profile.

**Full reciprocal-space mapping**

Synchrotron X-ray diffraction experiments were performed on the ID4B (QM2) beamline at the Cornell High Energy Synchrotron Source (CHESS). The incident X-ray energy was 50 keV ($\lambda$=0.248 Å), which was selected using a double-bounce diamond monochromator. An area detector (Pilatus 6M) was used to collect the scattering intensities in a reflection geometry. The sample was rotated with three tilted 360º rotations, sliced into 0.1º frames. Geometric parameters of the Pilatus6M detector such as detector distance, tilting, rotation, and direct beam position were extracted using standard $CeO_2$ powder, and data was processed and analyzed using in-house software. The data visualization was performed with the NeXpy software package.

**Micro-spectroscopic ellipsometry experiments**

Imaging spectroscopic ellipsometry was conducted to extract the dielectric function of perovskite using the Accurion EP4 system (Park Systems). Measurements were performed across the 250–1000 nm spectral range with a 7× objective lens. To ensure measurement reliability and accurately determine material properties, multi-angle incidence measurements were performed over an angular range of 40° to 60°. The ellipsometric coefficients maps ($\Psi$ and $\Delta$) were obtained with a spatial resolution of approximately 1 µm within the region of interest. The map data were analyzed using the EP4 model and DataStudio software provided with the EP4 system.

For ellipsometric coefficients analysis ($\Psi$ and $\Delta$), coefficients were extracted from the target flakes in map data using DataStudio software and subsequently analyzed using the EP4 model software. For modeling the excitonic behavior of perovskite, we employed the multi-Lorentz oscillator model as: $\varepsilon(E) = \varepsilon_\infty + \sum_i \frac{A_i}{E_{0,i}^2 - E^2 - i\Gamma_i E}$, where $\varepsilon_\infty$ is a background permittivity, $E_{0,i}$ denotes the resonance energy of the *i*-th oscillator, $A_i$ represents its oscillator strength, and $\Gamma_i$ is its damping factor. Based on the crystal structure of perovskite, the material was modeled as isotropic material. The parameters of each Lorentz oscillator and background permittivity were fitted to the experimental data to determine the complex refractive indices.

**Inelastic X-ray scattering**

The inelastic X-ray scattering (IXS) was conducted at beamline 10-ID at the NSLS-II [IXS].[62] The highly monochromatic x-ray beam of incident energy $E_i$ = 9.13 keV ($\lambda$ = 1.36 Å) was focused on the sample with a beam cross-section of ~10 × 10 μm$^2$. The total energy resolution was $\Delta E$ ~ 1.4 meV (full width at half maximum). The measurements were performed in reflection geometry on sample's (222) surface.

**DFT calculations**

For the lattice constant, DFT calculations were performed with VASP[63, 64, 65] using the projector-augmented wave (PAW) method.[66] A plane-wave cutoff energy of 400 eV and a Γ-centered 4×4×4 Monkhorst–Pack k-point mesh were employed. Structural relaxations used the PBEsol exchange-correlation functional[67, 68] until the residual forces on all atoms were less than 0.01 eV/Å.

To obtain the projected density of states (**Fig. S22**), DFT calculations were performed within the generalized-gradient approximation (GGA) of Perdew, Burke, and Ernzerhof (PBE)[67] as implemented in Quantum ESPRESSO[69], using a fully relativistic spinor formalism and norm-conserving PseudoDojo pseudopotentials[70] to account for spin–orbit coupling from heavy elements. To model defects, we constructed 2×2×2 supercells of CAIC (320 atoms) with and without defects. Self-consistent and non-self-consistent runs employed a 3×3×3 Monkhorst–Pack k-point grid and a plane-wave kinetic-energy cutoff of 50~Ry. PBE typically underestimates the bandgap; however, it is sufficiently reliable for qualitatively analyzing bandgap changes induced by structural variations.

**Data availability**

Data sets generated during the current study are available from the corresponding author on reasonable request.

**Acknowledgements**

The work at Yale was primarily supported by the Air Force Office of Scientific Research (Grant No. FA9550-22-1-0209). D.J. and B.C. acknowledge support from Office of Naval Research Young Investigator Award, Metamaterials Program N00014-23-1-203. Calculations of supercell DOS were supported by the National Science Foundation Division of Chemistry under award number CHE-2412412 (X.X., D.Y.Q.). The calculations used resources of the National Energy Research Scientific Computing (NERSC), a DOE Office of Science User Facility operated under contract no. DE-AC02-05CH11231, under award BES-ERCAP-0031507, BES-ERCAP-0027380, BES-ERCAP-0028897 and BES-ERCAP-003284; and the Texas Advanced Computing Center (TACC) at The University of Texas at Austin. This research used resources of the Advanced Photon Source; a U.S. Department of Energy (DOE) Office of Science User Facility operated for the DOE Office of Science by Argonne National Laboratory under Contract No. DE-AC02-06CH11357. Work performed at the Center for Nanoscale Materials, a U.S. Department of Energy Office of Science User Facility, was supported by the U.S. DOE, Office of Basic Energy Sciences, under Contract No. DE-AC02-06CH11357. This research used resources 10-ID of the National Synchrotron Light Source II, a U.S. Department of Energy (DOE) Office of Science User Facility operated for the DOE Office of Science by Brookhaven National Laboratory under Contract No. DE-SC0012704. This work is based on research conducted at the Center for High-Energy X-ray Sciences (CHEXS), which is supported by the National Science Foundation (BIO, ENG and MPS Directorates) under award DMR-2342336. We thank Hao-Wu Lin, Hao-Cheng Lin, Lihua Zhang, Yanyan Li for assistance with sample fabrication and testing, and Dr. Hua Zhou for insightful discussions.

**Author contributions**

P.G. conceived and supervised the project. S.L. synthesized the samples and performed optical experiments. S.L. performed TR-XRD, XAS, and IXC experiments with B.G., D.A.W., H.W., C.A.K, D.C., C.L., X.Z., Y.Q.C., P.G., S.S., Z.K., and Y.H. B.T.D. synthesized the nanocrystals samples. X.X., X.W., D.Y.Q. and Y.Y. performed first-principles calculations. B.C. and D.J. performed micro-spectroscopic ellipsometry experiments. B.L. assisted with sample fabrication. S.L. wrote the manuscript together with P.G. All authors provided input and comments on the manuscript.

**Competing interests:** The authors declare no competing interests.

**Correspondence** and requests for materials should be addressed to Peijun Guo.

# References

1. Herz LM. Charge-Carrier Dynamics in Organic-Inorganic Metal Halide Perovskites. *Annual Review of Physical Chemistry* **67**, 65-89 (2016).

2. Stranks SD, *et al.* Electron-Hole Diffusion Lengths Exceeding 1 Micrometer in an Organometal Trihalide Perovskite Absorber. *Science* **342**, 341-344 (2013).

3. Dong Q, *et al.* Electron-hole diffusion lengths > 175 μm in solution-grown $CH_3NH_3PbI_3$ single crystals. *Science* **347**, 967-970 (2015).

4. Filip MR, Eperon GE, Snaith HJ, Giustino F. Steric engineering of metal-halide perovskites with tunable optical band gaps. *Nature Communications* **5**, 5757 (2014).

5. Kojima A, Teshima K, Shirai Y, Miyasaka T. Organometal Halide Perovskites as Visible-Light Sensitizers for Photovoltaic Cells. *Journal of the American Chemical Society* **131**, 6050-6051 (2009).

6. Xiong Z, *et al.* Homogenized chlorine distribution for >27% power conversion efficiency in perovskite solar cells. *Science* **390**, 638-642 (2025).

7. Lin K, *et al.* Perovskite light-emitting diodes with external quantum efficiency exceeding 20 per cent. *Nature* **562**, 245-248 (2018).

8. Ke Y, *et al.* High performance tandem perovskite LEDs through interlayer photon recycling. *Nature*, (2025).

9. Tsarev S, *et al.* Vertically stacked monolithic perovskite colour photodetectors. *Nature* **642**, 592-598 (2025).

10. Saidaminov MI, *et al.* Planar-integrated single-crystalline perovskite photodetectors. *Nature Communications* **6**, 8724 (2015).

11. Lee H-D, *et al.* Valley-centre tandem perovskite light-emitting diodes. *Nature Nanotechnology* **19**, 624-631 (2024).

12. Jeong J, *et al.* Pseudo-halide anion engineering for α-$FAPbI_3$ perovskite solar cells. *Nature* **592**, 381-385 (2021).

13. Heo JH, Han HJ, Kim D, Ahn TK, Im SH. Hysteresis-less inverted $CH_3NH_3PbI_3$ planar perovskite hybrid solar cells with 18.1% power conversion efficiency. *Energy & Environmental Science* **8**, 1602-1608 (2015).

14. Zhu H, *et al.* Long-term operating stability in perovskite photovoltaics. *Nature Reviews Materials* **8**, 569-586 (2023).

15. Li N, Niu X, Chen Q, Zhou H. Towards commercialization: the operational stability of perovskite solar cells. *Chemical Society Reviews* **49**, 8235-8286 (2020).

16. Slavney AH, Hu T, Lindenberg AM, Karunadasa HI. A Bismuth-Halide Double Perovskite with Long Carrier Recombination Lifetime for Photovoltaic Applications. *Journal of the American Chemical Society* **138**, 2138-2141 (2016).

17. Volonakis G, *et al.* $Cs_2InAgCl_6$: A New Lead-Free Halide Double Perovskite with Direct Band Gap. *The Journal of Physical Chemistry Letters* **8**, 772-778 (2017).

18. Pan W, *et al.* $Cs_2AgBiBr_6$ single-crystal X-ray detectors with a low detection limit. *Nature Photonics* **11**, 726-732 (2017).

19. Chu L, *et al.* Lead-Free Halide Double Perovskite Materials: A New Superstar Toward Green and Stable Optoelectronic Applications. *Nano-Micro Letters* **11**, 16 (2019).

20. Du K-z, Meng W, Wang X, Yan Y, Mitzi DB. Bandgap Engineering of Lead-Free Double Perovskite $Cs_2AgBiBr_6$ through Trivalent Metal Alloying. *Angewandte Chemie International Edition* **56**, 8158-8162 (2017).

21. Luo J, *et al.* Efficient and stable emission of warm-white light from lead-free halide double perovskites. *Nature* **563**, 541-545 (2018).

22. Liu Y, Nag A, Manna L, Xia Z. Lead-Free Double Perovskite $Cs_2AgInCl_6$. *Angewandte Chemie International Edition* **60**, 11592-11603 (2021).

23. Klarbring J, Hellman O, Abrikosov IA, Simak SI. Anharmonicity and Ultralow Thermal Conductivity in Lead-Free Halide Double Perovskites. *Physical Review Letters* **125**, 045701 (2020).

24. Ning W, *et al.* Thermochromic Lead-Free Halide Double Perovskites. *Advanced Functional Materials* **29**, 1807375 (2019).

25. de Paula AM, *et al.* Time-Domain Observation of Ultrafast Self-Trapped Exciton Formation in Lead-Free Double Halide Perovskites. *Journal of the American Chemical Society* **147**, 28923-28931 (2025).

26. Li S, Luo J, Liu J, Tang J. Self-Trapped Excitons in All-Inorganic Halide Perovskites: Fundamentals, Status, and Potential Applications. *The Journal of Physical Chemistry Letters* **10**, 1999-2007 (2019).

27. Hu T, *et al.* Mechanism for Broadband White-Light Emission from Two-Dimensional (110) Hybrid Perovskites. *The Journal of Physical Chemistry Letters* **7**, 2258-2263 (2016).

28. Benin BM, *et al.* Highly Emissive Self-Trapped Excitons in Fully Inorganic Zero-Dimensional Tin Halides. *Angewandte Chemie International Edition* **57**, 11329-11333 (2018).

29. Paritmongkol W, Powers ER, Dahod NS, Tisdale WA. Two Origins of Broadband Emission in Multilayered 2D Lead Iodide Perovskites. *The Journal of Physical Chemistry Letters* **11**, 8565-8572 (2020).

30. Li X, *et al.* Terahertz field-induced ferroelectricity in quantum paraelectric $SrTiO_3$. *Science* **364**, 1079-1082 (2019).

31. Ilyas B, *et al.* Terahertz field-induced metastable magnetization near criticality in $FePS_3$. *Nature* **636**, 609-614 (2024).

32. Nova TF, Disa AS, Fechner M, Cavalleri A. Metastable ferroelectricity in optically strained $SrTiO_3$. *Science* **364**, 1075-1079 (2019).

33. Stoica VA, *et al.* Non-equilibrium pathways to emergent polar supertextures. *Nature Materials* **23**, 1394-1401 (2024).

34. Lei H, Hardy D, Gao F. Lead-Free Double Perovskite $Cs_2AgBiBr_6$: Fundamentals, Applications, and Perspectives. *Advanced Functional Materials* **31**, 2105898 (2021).

35. Thomaz JE, Lindquist KP, Karunadasa HI, Fayer MD. Single Ensemble Non-exponential Photoluminescent Population Decays from a Broadband White-Light-Emitting Perovskite. *Journal of the American Chemical Society* **142**, 16622-16631 (2020).

36. Laermer F, Israel W, Elsaesser T. Femtosecond spectroscopy of stimulated emission from highly excited dye molecules. *J Opt Soc Am B* **7**, 1604-1609 (1990).

37. Li S, Dai Z, Li L, Padture NP, Guo P. Time-resolved vibrational-pump visible-probe spectroscopy for thermal conductivity measurement of metal-halide perovskites. *Review of Scientific Instruments* **93**, 053003 (2022).

38. Guzelturk B, *et al.* Understanding and Controlling Photothermal Responses in MXenes. *Nano Letters* **23**, 2677-2686 (2023).

39. Li B, *et al.* Dual-Hyperspectral Optical Pump–Probe Microscopy with Single-Nanosecond Time Resolution. *Journal of the American Chemical Society* **146**, 2187-2195 (2024).

40. Yang J, Zhang P, Wei S-H. Band Structure Engineering of $Cs_2AgBiBr_6$ Perovskite through Order–Disordered Transition: A First-Principle Study. *The Journal of Physical Chemistry Letters* **9**, 31-35 (2018).

41. Yazdani N, *et al.* Coupling to octahedral tilts in halide perovskite nanocrystals induces phonon-mediated attractive interactions between excitons. *Nature Physics* **20**, 47-53 (2024).

42. Jiang H, Cui S, Chen Y, Zhong H. Ion exchange for halide perovskite: From nanocrystal to bulk materials. *Nano Select* **2**, 2040-2060 (2021).

43. Li G, Zhang T, Guo N, Xu F, Qian X, Zhao Y. Ion-Exchange-Induced 2D–3D Conversion of $HMA_{1-x}FA_xPbI_3Cl$ Perovskite into a High-Quality $MA_{1-x}FA_xPbI_3$ Perovskite. *Angewandte Chemie International Edition* **55**, 13460-13464 (2016).

44. Knight AJ, Herz LM. Preventing phase segregation in mixed-halide perovskites: a perspective. *Energy & Environmental Science* **13**, 2024-2046 (2020).

45. Slotcavage DJ, Karunadasa HI, McGehee MD. Light-Induced Phase Segregation in Halide-Perovskite Absorbers. *ACS Energy Letters* **1**, 1199-1205 (2016).

46. Tang X, *et al.* Local Observation of Phase Segregation in Mixed-Halide Perovskite. *Nano Letters* **18**, 2172-2178 (2018).

47. Li S, Dai Z, Kocoj CA, Altman EI, Padture NP, Guo P. Photothermally induced, reversible phase transition in methylammonium lead triiodide. *Matter* **6**, 460-474 (2023).

48. Li S, Li B, Xie B, Ma Y, Mohanraj S, Guo P. All-Optical Probing of Phase Transition Dynamics in 2D Lead Halide Perovskites. *ACS Photonics* **11**, 4507-4514 (2024).

49. Li Y, *et al.* Dynamic Local Order and Ultralow Thermal Conductivity of $Cs_2AgBiBr_6$. *Nano Letters* **25**, 401-409 (2025).

50. He X, *et al.* Multiple Lattice Instabilities and Complex Ground State in $Cs_2AgBiBr_6$. *PRX Energy* **3**, 013014 (2024).

51. Sulaiman KO, Sudheeshkumar V, Scott RWJ. Activation of atomically precise silver clusters on carbon supports for styrene oxidation reactions. *RSC Advances* **9**, 28019-28027 (2019).

52. Manuel D, Cabaret D, Brouder C, Sainctavit P, Bordage A, Trcera N. Experimental evidence of thermal fluctuations on the x-ray absorption near-edge structure at the aluminum K edge. *Physical Review B* **85**, 224108 (2012).

53. Li W, *et al.* Regulation of the order–disorder phase transition in a $Cs_2NaFeCl_6$ double perovskite towards reversible thermochromic application. *Journal of Semiconductors* **42**, 072202 (2021).

54. Lassoued MS, Wang T, Faizan A, Li Q-W, Chen W-P, Zheng Y-Z. Reversible thermochromism, temperature-dependent conductivity and high stability for a laminated bismuth(iii)–silver(i) hybrid double perovskite. *Journal of Materials Chemistry C* **10**, 12574-12581 (2022).

55. Liu Y, *et al.* Emerging Thermochromic Perovskite Materials: Insights into Fundamentals, Recent Advances and Applications. *Advanced Functional Materials* **34**, 2402234 (2024).

56. Xu K-X, *et al.* High-order Raman scattering mediated by self-trapped exciton in halide double perovskite. *Physical Review B* **106**, 085205 (2022).

57. Liang L, Wang C, Chen J, Wang QJ, Liu X. Incoherent broadband mid-infrared detection with lanthanide nanotransducers. *Nature Photonics* **16**, 712-717 (2022).

58. Gong X, *et al.* Heterojunction floating-gate phototransistors for ultraweak short-wavelength infrared photodetection. *Journal of Materials Chemistry C* **13**, 12762-12771 (2025).

59. Locardi F, *et al.* Emissive Bi-Doped Double Perovskite $Cs_2Ag_{1–x}Na_xInCl_6$ Nanocrystals. *ACS Energy Letters* **4**, 1976-1982 (2019).

60. Kinigstein ED, *et al.* X-ray multi-probe data acquisition: A novel technique for laser pump x-ray transient absorption spectroscopy. *Review of Scientific Instruments* **92**, (2021).

61. Kinigstein ED, *et al.* Asynchronous x-ray multiprobe data acquisition for x-ray transient absorption spectroscopy. *Review of Scientific Instruments* **94**, (2023).

62. Cai YQ, *et al.* The meV-resolved Inelastic X-ray Scattering Beamline at NSLS-II: Design and Performance. *Journal of Physics: Conference Series* **3010**, 012102 (2025).

63. Kresse G, Hafner J. Ab initio molecular dynamics for liquid metals. *Physical Review B* **47**, 558-561 (1993).

64. Kresse G, Hafner J. Ab initio molecular-dynamics simulation of the liquid-metal--amorphous-semiconductor transition in germanium. *Physical Review B* **49**, 14251-14269 (1994).

65. Kresse G, Furthmüller J. Efficiency of ab-initio total energy calculations for metals and semiconductors using a plane-wave basis set. *Computational Materials Science* **6**, 15-50 (1996).

66. Blöchl PE. Projector augmented-wave method. *Physical Review B* **50**, 17953-17979 (1994).

67. Perdew JP, Burke K, Ernzerhof M. Generalized Gradient Approximation Made Simple. *Physical Review Letters* **77**, 3865-3868 (1996).

68. Perdew JP, *et al.* Restoring the Density-Gradient Expansion for Exchange in Solids and Surfaces. *Physical Review Letters* **100**, 136406 (2008).

69. Giannozzi P, *et al.* QUANTUM ESPRESSO: a modular and open-source software project for quantum simulations of materials. *Journal of Physics: Condensed Matter* **21**, 395502 (2009).

70. Hamann DR. Optimized norm-conserving Vanderbilt pseudopotentials. *Physical Review B* **88**, 085117 (2013).